# COMPUTATION OF THE SHOCK ADIABATS OF ARGON AND XENON

V. K. Gryaznov, I. L. Iosilevskii, and V. E. Formov

UDC 533.951:533.6.011.72.001.24

Results of computations of the shock adiabats of Ar and Xe taking account of ionization, electron excitation, and a nonideal plasma are represented. Energy losses by radiation are estimated.

The domain of parameters accessible for a dynamic experiment can be broadened substantially [1] by recording the states orginating during expansion of a shock-compressed material in a medium ("obstacle") with a lesser dynamic impedance and known thermodynamic properties. Gases [2, 3] whose dynamic impedance is varied easily by changing the initial pressure and the molecular weight are hence used as obstacles for measurements in the "low" pressure domain ($P \lesssim 50$ kbar).

Results are presented herein of computations of the shock adiabats of argon and xenon in a range of parameters of interest for experiments to record the unloading isentrope of inert condensed media and explosion products. The shock adiabats of Ar and Xe are also of interest in connection with research to produce explosive sources of a luminous plasma [4, 5]. The shock adiabat calculations now available in the literature [6, 7] correspond to conditions in shock tubes and refer essentially to lower initial pressures; data are presented in [8] without taking account of the contribution of bound states, which is significant under the conditions being considered.

During propagation of a stationary shock discontinuity, conservation laws which can be written

$$D = V_0 \left( \frac{P - P_0}{V_0 - V} \right)^{1/2} \tag{1}$$

$$u = [(P - P_0)(V_0 - V)]^{1/2} \tag{2}$$

$$H - H_0 = \tfrac{1}{2}(P - P_0)(V_0 + V) \tag{3}$$

when radiation transfer is neglected, are satisfied on its front.

Here D, u are the velocity of the shock and the shock-compressed material in the laboratory coordinate system; H, V, P are the specific enthalpy, the volume, and the pressure; and the subscript 0 on these parameters refers to the state in front of the shock discontinuity.

The system (1)-(3) is supplemented by the equation of state of the medium behind the shock front. In the range of parameters considered here, the shock-compressed gas is in the plasma state so that the electron excitation, ionization, and corrections to the thermodynamic functions because of Coulomb interaction must be taken into account in the thermodynamic computations. In conformity with [9], the equation of state of a multicomponent plasma is

$$\frac{P}{kT} = \sum_j n_j - \frac{\varkappa^3}{24\pi} \tag{4}$$

$$\frac{H}{VkT} = \frac{5}{2}\sum_j n_j - \frac{\varkappa^3}{6\pi} + \sum_j n_j \left[ J_j + \frac{1}{Q_j}\sum_m g_m^j \exp\left(-\frac{E_m^j}{kT}\right) \right] \tag{5}$$

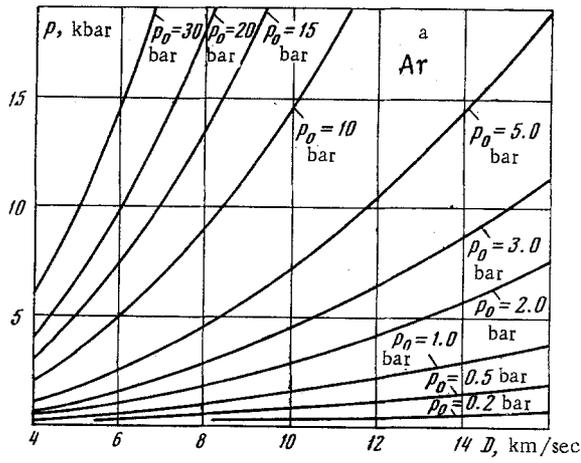
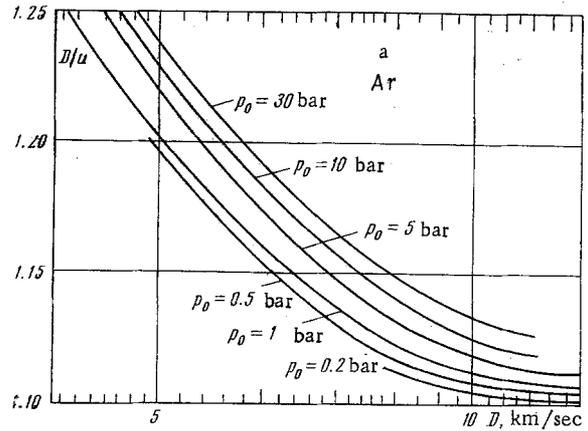
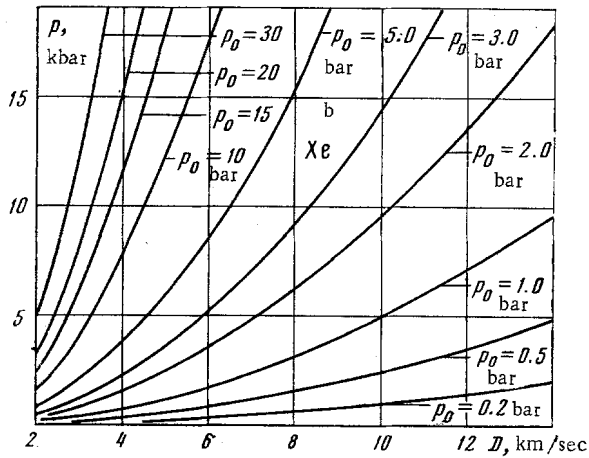
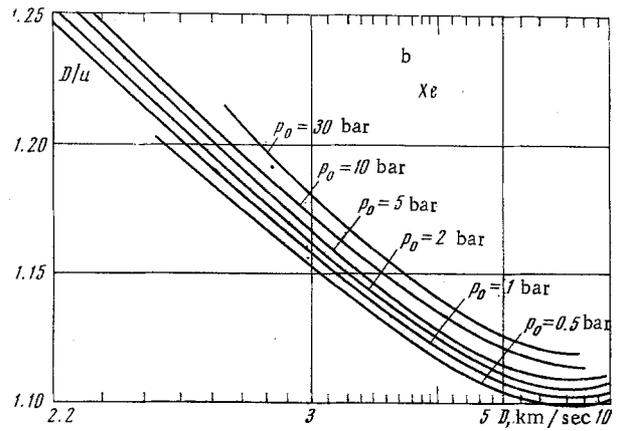

Fig. 1                    Fig. 2

The ionization equilibrium equations (Saha formulas) are

$$\frac{n_j n_e}{n_{j-1}} = \frac{2Q_j}{Q_{j-1}} \lambda_e^{-3} \exp\left(-\frac{J_j - \chi_j}{kT}\right) \quad (\lambda_e = [h^2/2mkT]^{1/2}) \tag{6}$$

The reduction in ionization potential caused by the Coulomb interaction is

$$\chi_j = \ln \frac{[1+\gamma/2][1+z_j^2\gamma/2]}{[1+(z_j-1)^2\gamma/2]} \tag{7}$$

Here $\lambda_e$ is the thermal De Broglie wavelength of the electron, $\varkappa$ and $\gamma$ are connected by the relationship $\gamma = \varkappa e^2/kT$, $\gamma$ is a positive root of the equation

$$\gamma^2 = \left(\frac{e^2}{kT}\right)^3 \left[4\pi \sum_k n_k z_k^2 \Big/ \left(1 + z_k^2 \frac{\gamma}{2}\right)\right]$$

$n_j$, $z_j$, $J_j$ are the density, charge, and ionization potential of particles of the species j. The subcript j in (4)-(7) enumerates all the particles, and k numbers just the charged particles.

In the weak nonideal case, the parameter $\gamma$ agrees with the usual nonideal parameter

$$\Gamma = (e^2/kT)^{3/2} (4\pi \sum_k n_k Z_k^2)^{1/2}$$

The statistical sum of the j-th particle

$$Q_j = \sum_m g_m^j \exp(-E_m^j/kT) \tag{8}$$

is evaluated by means of the tabulated values [10] of the statistical weights $g_m^j$ and the excitation energy $E_m^j$ of the m-th energy level. Terms with the energies $E_m^j < J_j - \chi_j$ were taken into account in evaluating $Q_j$ in (8). The corrections to the thermodynamic functions because of Coulomb interaction of the free charges in (4)-(7) are written in conformity with the Debye theory in a large canonical ensemble of statistical mechanics [9].

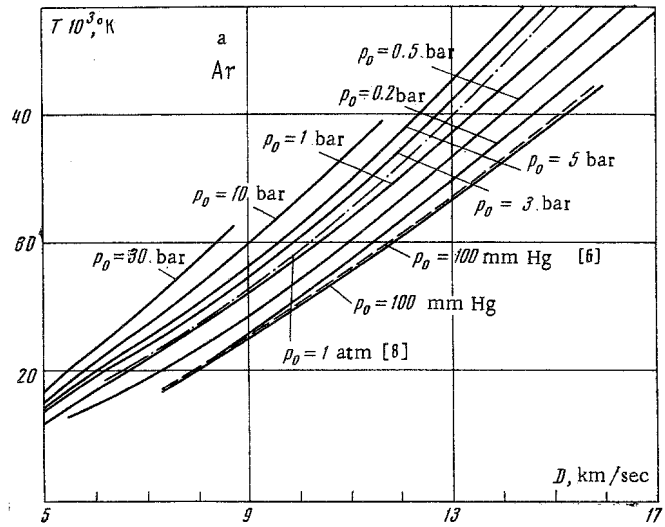

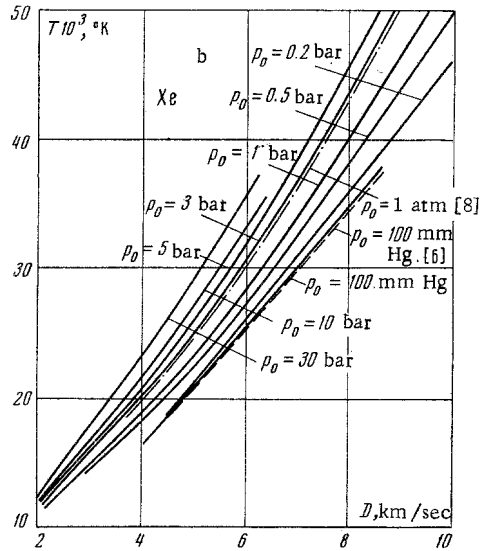

Fig. 3

The corrections for a nonideal plasma given by this theory for small interparticle interaction ($\Gamma \ll 1$) agree with the customary Debye approximation and possess good (experimentally established [11]) extrapolation properties in the domain of elevated imperfection.

The closed system (1)-(8) was separated into two parts, hydrodynamic (1)-(3) and thermodynamic (4)-(8), for solution on a computer. The thermodynamic computation was carried out by a special program which found the equilibrium composition and thermodynamic functions of a multiply ionized gas for a given temperature and pressure. To do this, the equation of state (4) and the ionization equilibrium equations (6) were supplemented by the condition of electrical neutrality

$$\sum_k z_k n_k = 0 \qquad (9)$$

and the system of nonlinear equations in $n_k$ obtained was solved in two iteration cycles. The corrections for imperfection in (4), (6) were considered constant in the inner cycle, and the system of equations was hence solved by the gradient descent method. The composition obtained in this manner was used in the outer cycle to convert the corrections for imperfection $\Delta P$, $\Delta H$, and $\chi_j$, after which the composition was again found, etc., until the complete convergence of the concentrations of all the components. It should be noted that the form selected for the corrections to imperfection (4)-(7) assured the mentioned convergence of the system even in the case of a strongly nonideal ($\Gamma \gtrsim 1$) plasma.

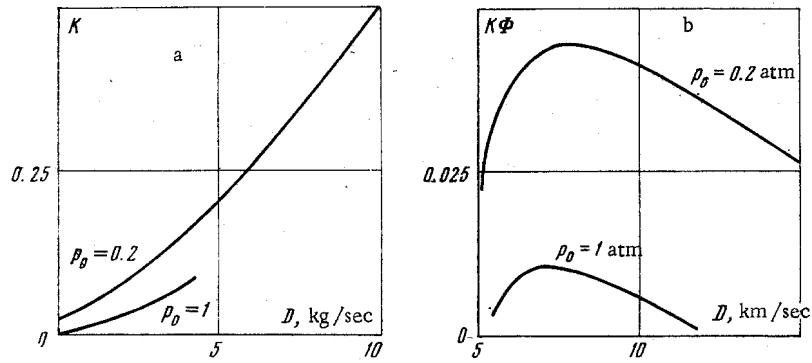

Fig. 4

The results of the computations are presented in Figs. 1-3. The graphs are given for diverse values of the initial gas pressure $P_0$ as a function of the shock front velocity D, quantities measured most easily and accurately in experiments. In conformity with the condition of continuity of the pressure P and the mass flow rate u on the interface (the contact discontinuity) between the material under investigation and the shock-compressed obstacle [12], the graphs in Figs. 1 and 2 permit finding the state in an isentropic unloading wave on the P−u plane by measuring the shock velocity D in the obstacle.

The computed values of the temperature of the shock-compressed plasma are presented in Fig. 3 and are compared (dashes) with the results in [8], where the contribution of the bound states was not taken into account in the Saha equation and the enthalpy. This comparison, as well as the results of calculations carried out taking account of just the ground energy state of the atoms and ions in (5), (6), show that the contribution of the bound states in the thermodynamic functions of a shock-compressed plasma increases as the initial pressure $P_0$ rises and is substantial in the range of parameters under consideration here. The states of Ar, Xe plasma computed by such a method for $P_0 = 0.132$ atm agree with the data in [6]; negligible (3% in T) deviations are caused by the differences in taking account of the plasma imperfection and the contribution of the bound states in (5) and (6).

The influence of the plasma imperfection on the parameters behind the shock front increases analogously to the contribution of the bound states as the initial pressure $P_0$ rises. Computations of the shock adiabats of Ar, Xe made in an ideal gas approximation show that, at high shock velocities and initial pressures, taking account of the plasma imperfection increases the degree of compression ∼10% and decreases the temperature corresponding to the same shock velocity by ∼7%. Therefore, taking account of the influence of plasma imperfection in the range of parameters under consideration is just as substantial.

Computations show (Fig. 3) that high temperatures are developed behind the shock front so that the radiation flux issuing from the shock front [12] can be great under these conditions. Considering the thermal radiation to be in local thermodynamic equilibrium with the substance, and the shock front to give light as a blackbody,* let us write the ratio between the radiant energy flux and the hydrodynamic flux as $K = \sigma T^4 V/D$ (H−PV) ($\sigma$ is the Stefan−Boltzmann constant). It is seen from the results of computing this quantity (Fig. 4a) that at low initial pressures $P_0$ of xenon, the flux of light radiation can reach half the hydrodynamic value; for argon $K < 0.05$ in the range of parameters examined here.

A major part of the energy radiated by the front is, however, shielded by the cold gas ahead of the shock front [12]. The fraction of radiation energy arriving at the domain of gas transparency from $\lambda = \infty$ to the boundary of the ionization continuum $\lambda' \sim 0.1\,\mu$ and thence departing towards infinity is given by the Planck formula

$$\psi = \int_{\lambda'}^{\infty} a_1 \lambda^{-5} d\lambda \left( \exp\left[\frac{a_2}{\lambda T}\right] - 1 \right) \left[ \int_0^{\infty} a_1 \lambda^{-5} d\lambda \left( \exp\left[\frac{a_2}{\lambda T}\right] - 1 \right) \right]^{-1}$$

where $a_1, a_2$ are known constants [12].

---

*Results of estimates and experiments to produce a plasma by an explosion [8] are a sufficient basis for this.

The energy flux entrained by the radiation $K\psi$ turns out to be small (Fig. 4a), which affords the possibility of not taking it into account in the Hugoniot equation (3).